\documentclass[conference]{IEEEtran}
\IEEEoverridecommandlockouts
\usepackage{cite}
\usepackage{amsmath,amssymb,amsfonts}
\usepackage{algorithmic}
\usepackage{graphicx}
\usepackage{textcomp}
\usepackage{xcolor}
\usepackage{soul}
\usepackage{multirow}
\usepackage{array}
\usepackage{comment}
\usepackage{threeparttable}
\usepackage{fancyhdr}
\usepackage{hyperref}

\fancypagestyle{plain}{
  \fancyhf{}
  \fancyhead[C]{Conference on \LaTeX}     %% C or L or R.
  \fancyfoot[L]{This is a notice}%                        %% C or L or R.

}
\usepackage{eso-pic}

\def\BibTeX{{\rm B\kern-.05em{\sc i\kern-.025em b}\kern-.08em
    T\kern-.1667em\lower.7ex\hbox{E}\kern-.125emX}}

\makeatletter
\newcommand{\linebreakand}{%
  \end{@IEEEauthorhalign}
  \hfill\mbox{}\par
  \mbox{}\hfill\begin{@IEEEauthorhalign}
}
\makeatother

\begin{document}
\bstctlcite{IEEEexample:BSTcontrol}
\title{On the Echogenicity of Natural Starch-Based Blood Mimicking Fluids for Contrast Enhanced Ultrasound Imaging: Preliminary \textit{In-vitro} Experiments\\
\thanks{$^{\dagger}$Equal Contributions}
}

\author{V. Arun Kumar$^{1\dagger}$, A. N. Madhavanunni$^{2\dagger}$, S. Nivetha$^{3}$, Mahesh Raveendranatha Panicker$^{4}$\\%} %\\ \\
\textit{$^{1,2}$Indian Institute of Technology Palakkad, Kerala, India}\\
      \textit{ $^{3}$Vivekanandha College of Engineering for Women, Namakkal, Tamil Nadu, India}\\
      \textit{ $^{4}$Singapore Institute of Technology, Singapore}\\
      \small $^{1}$122001049@smail.iitpkd.ac.in, $^{2}$121813001@smail.iitpkd.ac.in, $^{3}$s.nivetha7845@gmail.com, $^{4}$mahesh.panicker@singaporetech.edu.sg}

% \author{\IEEEauthorblockN{V. Arun Kumar}
% \IEEEauthorblockA{\textit{Dept. of Electrical Engineering} \\
% \textit{Indian Institute of Technology Palakkad}\\
% Palakkad, Kerala, India \\
% 122001049@smail.iitpkd.ac.in}
% \and 
% \IEEEauthorblockN{A. N. Madhavanunni}
% \IEEEauthorblockA{\textit{Dept. of Electrical Engineering} \\
% \textit{Indian Institute of Technology Palakkad}\\
% Palakkad, Kerala, India \\
% 121813001@smail.iitpkd.ac.in}
% \linebreakand
% \IEEEauthorblockN{S. Nivetha}
% \IEEEauthorblockA{\textit{Dept. of Biotechnology} \\
% \textit{Vivekanandha College of Engg. for Women}\\
% Namakkal, Tamil Nadu, India \\
% s.nivetha7845@gmail.com}
% \and
% \IEEEauthorblockN{Mahesh Raveendranatha Panicker}
% \IEEEauthorblockA{\textit{Infocomm Technology Cluster} \\
% \textit{Singapore Institute of Technology}\\
% Singapore \\
% mahesh.panicker@singaporetech.edu.sg}
% }

\begin{titlepage}
    %\centering
    \vspace*{\fill}
\fontsize{15}{18}\selectfont\textcopyright { 2024 IEEE. Personal use of this material is permitted. Permission from IEEE must be obtained for all other uses, in any current or future media, including reprinting/republishing this material for advertising or promotional purposes, creating new collective works, for resale or redistribution to servers or lists, or reuse of any copyrighted component of this work in other works.}
    \vspace*{\fill}
\end{titlepage}

\AddToShipoutPictureBG{%
  \AtPageUpperLeft{%
    \setlength\unitlength{1in}%
    \hspace*{\dimexpr0.5\paperwidth\relax}%%  change \dimexpr0.5\paperwidth\relax appropriately
    \makebox(0,-0.75)[c]{{This work has been accepted in the IEEE South Asian Ultrasonics Symposium 2024 (IEEE SAUS 2024).}}%

    % \makebox(0,-0.75)[c]{\small {This is an originally submitted version and has not been reviewed by independent peers.}}%
    % {\large This is an originally submitted version and has not been reviewed by independent peers.}}%
}}

\maketitle

\begin{abstract}
Natural starch-based blood-mimicking fluid (BMF) has been used as an alternative to commercially available BMFs for \textit{in-vitro} Doppler investigations in low-resource settings. Most reported works in the literature have used corn starch-based BMF. Evaluation of other natural starches for potential BMF and their characterization have relatively been unexplored in the literature. To this end, this work investigates the echogenicity of corn-, potato-, tapioca-, and wheat starch-based BMFs prepared using a liquid base of pure water-glycerol mixture with three different starch concentrations (1\%, 3\%, and 5\%). The experiments were performed by manually pumping the BMFs to a PolyVinyl alcohol (PVA) based flow phantom using a syringe and raw datasets were acquired using a Verasonics Vantage 128 Research Ultrasound System. Echogenicity was measured as the mean pixel intensity in a selected region of interest (ROI) in the beamformed image. Among the four natural starch-based BMFs, potato starch-based BMF showed the highest echogenicity and contrast with almost 13\%, 14\%, and 10\% higher pixel intensities (dB) than that of the least echoic BMF with 1\%, 3\%, and 5\% starch concentrations respectively. Moreover, the echogenicity of corn, tapioca, and wheat starch-based BMF was observed to be similar, and the results suggest that these BMFs with higher starch concentrations shall be employed for \textit{in-vitro} contrast-enhanced ultrasound imaging.
% Ultrasound imaging is a well-established clinical imaging technique that provides real-time, quantitative, anatomical, and physiological information in humans. However, owing to the anechoic nature of human blood, visualizing blood flow and human microvascular structures remains challenging with ultrasound. 
% %\st{To address this challenge, contrast agents (CAs) are available for contrast-enhanced ultrasound (CEUS) imaging. However, the cost and accessibility of commercially available CAs can be prohibitive in low-resource settings.}
% In this regard, to enhance the visualization of microvascular structures, contrast-enhanced ultrasound(CEUS) imaging is employed in studies. This study investigates the potential of readily available and inexpensive natural starches, such as cornstarch, potato starch, wheat starch, and tapioca starch, as blood mimicking fluids(BMFs) for in-vitro CEUS imaging.. By comparing the echogenicity of these four natural starch-based BMFs, we demonstrate the potential of these natural alternatives for affordable and accessible \textit{in- vitro} blood flow imaging in low-resource settings.
\end{abstract}
\begin{IEEEkeywords}
Blood-mimicking fluid, Contrast-enhanced ultrasound imaging, Echogenicity, Natural starch
\end{IEEEkeywords}

% However, many of the \textit{in-vitro} investigations rely on custom-prepared starch-based blood-mimicking fluid (BMF) instead of commercial BMFs. 
\begin{comment} 

\textcolor{blue}{Natural starch-based blood-mimicking fluid (BMF) has been used as an alternative to commercially available BMFs for in-vitro Doppler investigations in low-resource settings. Most reported works in the literature have used corn starch-based BMF. Evaluation of other natural starches for potential BMF and their characterization have relatively been unexplored in the literature. In this regard, this paper investigates the echogenicity of four natural starch-based BMFs for in-vitro contrast-enhanced ultrasound (CEUS) imaging.}

\end{comment}

\section{Introduction} \label{sec:intro}
Ultrasound (US) imaging is a well-established clinical imaging technique that provides real-time, quantitative anatomical and physiological information in humans. The lack of ionizing radiation, the ability to do dynamic imaging, and relatively low purchase and maintenance costs make it the most affordable clinical imaging technique with increasing use for guiding interventional clinical procedures. 
%US imaging is considered the gold standard for noninvasive blood flow imaging. 
% However, despite its widespread use and numerous advantages, conventional ultrasound imaging does come with some limitations. For instance, blood flow imaging may be challenging since human blood is typically anechoic or hypoechoic, making it less visible during ultrasound imaging. 
Even though US imaging is considered the gold standard for noninvasive blood flow imaging, visualizing blood flow and human microvascular structures remains challenging because of the anechoic nature of human blood. To address this and enhance the visualization of microvascular structures, contrast-enhanced ultrasound (CEUS) imaging is employed, by administering US contrast agents intravenously to create synthetic scattering inside the blood vessels \cite{greis2009ultrasound,liu2008evaluation}.
% Contrast agents (CAs) play a major role in CEUS by increasing the echogenicity of blood, allowing for better differentiation of vascular structures and improved imaging quality. There have been many studies employing CEUS imaging, especially in enhancing the visibility of blood flow and vascular structures.\cite{mehta2017vascular,greis2011quantitative,broumas2005contrast}.  
% Numerous studies have explored the use of commercially available ultrasound contrast agents\cite{paefgen2015evolution, wei2001quantification}. 
Clinically made microbubbles are the commonly used contrast agents, composed of an outer shell covering a gaseous core in which the shell is typically composed of galactose, polymers, albumin, or lipids \cite{mehta2017vascular} for stabilizing the microbubbles\cite{koczera2017pbca}. 
These microbubbles have been used as contrast agents (CAs) for CEUS in detecting malignancies\cite{xu2009contrast} in the breast\cite{liu2008evaluation}, gallbladder\cite{xie2010differential}, bile duct\cite{xu2006imaging}, pancreas\cite{kersting2009quantitative}, kidney\cite{xu2010renal} etc.

However, to facilitate the development of novel ultrasonic imaging techniques for such studies, and to assess blood flow dynamics and related parameters \textit{in-vitro}, phantom test beds with known flow dynamics and parameters are necessary. Typically, those \textit{in-vitro} experimental setups consist of a tissue-mimicking phantom that mimics the acoustic property of tissues, a blood-mimicking fluid (BMF), CAs (for CEUS studies), and a controlled pumping mechanism to simulate the blood flow. %For CEUS imaging, are also pumped along with the BMF to enhance the echogenicity. 

%{However, a drawback for these contrast agents (CAs) is the cost and availability, especially in a low-resource setup. These challenges have led researchers to explore alternative CAs that are not only cost-effective but also readily available.} 

%Viscosity \cite{chai,reashini}, density\cite{chai,leibuss(Concentration not mentioned),attenuation\cite{secomski}}
Apart from the commercially available BMFs, custom-made BMFs were also used in the previously reported flow imaging studies \textit{in-vitro} \cite{yoshida2012blood,tanaka2012blood,oglat2018review}. Natural starch particles, particularly corn starch, have been the most popular choice to mimic blood cells and generate echoes from the flow channels due to their acoustic properties \cite{kargel2003adaptive, duerk1992physiologic}. Previous attempts in the acoustic characterization of the cornstarch-based BMFs have shown that their viscosity, density, and acoustic scattering are similar to that of blood \cite{chai2023blood, reashini2023effect, leibuss2022transcranial, secomski2017influence}.
%\textcolor{red}{Several studies have characterized these cornstarch-based BMFs to assess their similarity with blood. Specifically, the viscosity of BMFs has been tested by \cite{chai2023blood,reashini2023effect}, density by \cite{leibuss2022transcranial,reashini2023effect}, and acoustic scattering by \cite{secomski2017influence}.} 
% Corn starch have emerged as scattering fluid\cite{kargel2003adaptive}, reflective material\cite{duerk1992physiologic} for BMFS in ultrasound imaging for\textit{in-vitro} setups.
Moreover, corn starch-based BMFs were employed in numerous \textit{in-vitro} studies demonstrating color Doppler \cite{kargel2003adaptive, duerk1992physiologic}, spectral Doppler \cite{secomski2017influence}, power Doppler mapping \cite{lecart2009vitro, li2015coherent, ozgun2019spatial}, transducer artefact detection with B-mode imaging \cite{king2010evaluation}, and \textit{in-vitro} study of blood perfusion in the foot\cite{nair2022ultrafast}. 
% Angiography imaging \cite{chai2023blood, reashini2023effect}.
% \textcolor{red}{Arun to mention different studies/applications and corresponding references accordingly}.
% Clutter suppression in color flow imaging\cite{kargel2003adaptive},color doppler ultrasonics mapping\cite{duerk1992physiologic}, Angiography imaging\cite{chai2023blood,reashini2023effect},3D B-mode\cite{lecart2009vitro}, power doppler\cite{lecart2009vitro},artefact detection for any transducer model using B-mode imaging\cite{king2010evaluation},coherent flow power doppler (CFPD)\cite{li2015coherent,ozgun2019spatial},velocity of acoustic
% streaming from doppler spectrum\cite{secomski2017influence}, \textit{in-vitro} study of blood perfusion in the foot\cite{nair2022ultrafast}.\cite{chai2023blood,reashini2023effect,duerk1992physiologic,kargel2003adaptive,lecart2009vitro,king2010evaluation,li2015coherent,ozgun2019spatial,secomski2017influence,nair2022ultrafast} for \textit{in-vitro} studies, because of their biocompatibility, low cost, and ease of availability.
%demonstrating their potential to address the limitations of CEUS for \textit{in-vitro} setups. 
A detailed analysis of the related approaches is summarized in Table \ref{literature}. However, the evaluation of other natural starches for potential BMFs and their characterization has relatively been unexplored in the literature. In this regard, this study aims to investigate the echogenicity of four different naturally available starch-based BMFs for a potential replacement of commercial BMFs to be used for \textit{in-vitro} studies in low-resource settings.
%\textcolor{red}{Insert a table on the same}
% https://iopscience.iop.org/article/10.1088/1742-6596/2222/1/012016/pdf
% https://iopscience.iop.org/article/10.1143/JJAP.51.07GF18
% https://ieeexplore.ieee.org/stamp/stamp.jsp?tp=&arnumber=6561998
% https://www.ncbi.nlm.nih.gov/pmc/articles/PMC9944827/pdf/JMU-30-251.pdf
\begin{table*}[t]%[width=0.75\textwidth]
\begin{threeparttable}
\centering
\caption{Analysis of starch-based BMFs reported in the literature}\label{tbl:Litreview}
%\begin{tabular}{llll}
\begin{tabular*}{0.99\textwidth}{llll}
%\toprule
\hline \hline 
% \textbf{Source} & \textbf{Ingredients} & \textbf{Starch Concentration} & \textbf{Comments} \\
%  &  &  (wt/wt\%) &   \\
% \hline
\multirow{2}{3cm}{ \textbf{Source}}& \multirow{2}{5cm}{\textbf{Ingredients} } & \multirow{2}{3cm}{\textbf{Starch Concentration (wt/wt) }}  &  \multirow{2}{5cm}{\textbf{Comments}}\\\\     
\hline
\hline
\multirow{2}{3cm}{C. Kargel \textit{et al.}\cite{kargel2003adaptive}}& \multirow{2}{5cm}{Water, Corn Starch} &\multirow{2}{3cm}{1\%}  &  \multirow{2}{5.5cm}{Used for Color Flow imaging.
}\\\\ 
\hline
\multirow{4}{3cm}{J. L. Duerk \textit{et al.} \cite{duerk1992physiologic}}& \multirow{4}{5cm}{Water, Glycerol, Corn Starch} & \multirow{4}{3cm}{N.A\tnote{*}} &  \multirow{4}{5.5cm}{Buoyancy of the particles was increased by heating to 100$^{\circ}$ for several minutes and cooling to room temperature. Used in color Doppler ultrasound study.
}\\\\\\\\ 

\hline

\multirow{2}{3cm}{W. Secomski \textit{et al.} \cite{secomski2017influence}}& \multirow{2}{5cm}{Water, Corn Starch} & \multirow{2}{3cm}{0.1-0.4\%} &  \multirow{2}{5.5cm}{The streaming velocity increased as the concentration of CS increased.}\\\\ 
\hline

\multirow{2}{3cm}{M. Lécart \textit{et al.} \cite{lecart2009vitro}}& \multirow{2}{5cm}{Water, Glycerol, Corn Starch, Salt} & \multirow{2}{3cm}{17.48\% (v/v)} &  \multirow{2}{5.5cm}{Used for power Doppler recordings.
}\\\\ 
\hline

\multirow{4}{3cm}{Y. L. Li \textit{et al.} \cite{li2015coherent}}& \multirow{4}{5cm}{Water, Corn Starch} & \multirow{4}{3cm}{3\%} &  \multirow{4}{5.5cm}{Used to study the feasibility of flow detection using the spatial coherence of backscattered ultrasound in power Doppler imaging.}\\\\ \\\\
\hline
\multirow{2}{3cm}{K. Ozgun \textit{et al.}\cite{ozgun2019spatial}}& \multirow{2}{5cm}{Water, Corn Starch} & \multirow{2}{3cm}{0-3\%} &  \multirow{2}{5.5cm}{- %We propose the application of mutual intensity, rather than normalized coherence,to retain the clutter suppression capability
%inherent in coherence beamforming, while preserving the
%underlying signal energy
}\\\\ 

\hline
\multirow{4}{3cm}{D. M. King \textit{et al .}\cite{king2010evaluation}}& \multirow{4}{5cm}{Water, Corn Starch} & \multirow{4}{3cm}{0.1-6\%} &  \multirow{4}{5.5cm}{Corn starch synergized the interactions with xantham gum (XG), increasing the viscosity. Above 4\% concentration, the cornstarch is prone to clumping.}\\\\ \\\\
\hline
\multirow{2}{3cm}{H. S. Nair \textit{et al.}\cite{nair2022ultrafast}}& \multirow{2}{5cm}{Water, Glycerol Corn Starch} & \multirow{2}{3cm}{1\% (w/v)} &  \multirow{2}{5.5cm}{Used for ultrafast Doppler ultrasound in assessing the blood perfusion in the foot.
}\\\\ 
\hline
\multirow{3}{3cm}{S. Chai \textit{et al.} \cite{chai2023blood}}& \multirow{3}{5cm}{Water, Glycerol, Xantham Gum, Iohexol (CA), Surfactant, Corn Starch} & \multirow{3}{3cm}{0.01\%}  &  \multirow{3}{5.5cm}{Corn starch synergized the interactions with XG, increasing the viscosity. Used in angiography imaging.}\\\\\\    
\hline
\multirow{3}{3cm}{S. Reashini \textit{et al.}\cite{reashini2023effect}}& \multirow{3}{5cm}{Water, Glycerol, Xantham Gum, Iohexol 350, Surfactant, Corn Starch} &\multirow{3}{3cm}{0.01\%}  &  \multirow{3}{5.5cm}{Suspensions of CS and water exhibits shear thickening at shear rate  $> 1s^{-1}$. Used in angiography imaging.}\\\\\\
\hline
\multirow{3}{3cm}{N. Perrira \textit{et al.}\cite{perrira2022experimental}}& \multirow{3}{5cm}{Water, Glycerol, Xantham Gum, Iohexol, Surfactant, Starch} & \multirow{3}{3cm}{0-0.06\% (w/v)} &  \multirow{3}{5.5cm}{Addition of starch made the BMF exhibit shear thickening at low shear rates. Used in angiography imaging.
}\\\\\\
\hline \hline
\end{tabular*}
\begin{tablenotes}
            \small
            \item[*]\textit{Concentration not specified in the research}
        \end{tablenotes}
\end{threeparttable}
\label{literature}
\end{table*}

%and also to compare these starch-based BMFs with a commercially available BMF.
The rest of the paper is organized as follows. Section \ref{sec:methods} describes the BMF preparation, the experimental setup used for imaging, and the method employed for echogenicity measurement. Section \ref{sec:results} presents the B-mode and power Doppler images obtained \textit{in-vitro} for different BMFs and discusses their echogenicity quantitatively. 
% Section \ref{sec:conclusion} concludes the paper.

% Section \ref{sec:results} presents the B-mode images of PVA-based flow phantoms with different starch concentrations and discusses their echogenicity quantitatively. 

\section{Materials and Methods} \label{sec:methods}

\subsection{Blood Mimicking Fluid Preparation}

In this study, four distinct starch solutions were prepared using starch derived from corn, tapioca, potato, and wheat (Urban Platter \textregistered), with three different starch concentrations of $1\%$, $3\%$, and $5\%$. The solutions were prepared by mixing starch powder with a liquid base of pure water-glycerol mixture. The volume-to-volume ratio for the liquid base was adjusted such that a constant overall fluid volume was maintained, with water - glycerol ratios of $55\%-44\%$ for $1\%$ scratch, $53\%-44\%$ for $3\%$ scratch, and $51\%-44\%$ for $5\%$ starch concentrations.

\subsection{\textit{In-vitro} Experimental Setup}

\subsubsection{Tissue Mimicking Phantom}
The \textit{in-vitro} evaluation of natural starch-based BMFs was performed using a custom-made tissue-mimicking wall-less flow phantom. The phantom was designed with a straight wall-less channel of $\approx3\ mm$ diameter at a depth of $2.25 cm$ from the top surface. The phantom was prepared using $10\%$ (w/v) Poly Vinyl Alcohol (PVA, Elvanol 71-30), and $90\%$ (w/v) Deionized (DI) water according to a previously reported protocol \cite{mercado2018effect}. 
%and 1\% (w/v) Silicone Carbide. DI water served as the base solvent, and SiC particles mimicked tissue backscattering characteristics. 
A two-step freeze-thaw cycle (24 hrs) was employed to ensure the complete cross-linking of the PVA \cite{nair2022ultrafast}. 
% The phantom was designed in such a way that a single straight vessel with a diameter of approximately 3 mm was present at a depth of 2.25 cm from the surface. 
Since the objective of this study was to measure the echogenicity of the natural starch-based BMFs, no particles were added during phantom preparation for random scattering. The hypothesis was that this would help minimize the scattering signal other than from inside of the flow channel and enable easier interpretation of the B-modes qualitatively and quantitatively.
%\textcolor{red}{cite the papers on phantom preparation (Check the protocol document for the citation info)}

%%%%%%%
\begin{table}[t]%[width=0.75\textwidth,cols=3,pos=h]
\centering
\caption{Data Acquisition Parameters}\label{tbl:systemDetails}
%\begin{tabular}{llll}
\begin{tabular*}{0.4\textwidth}{ll}
%\toprule
\hline \hline 
Parameter         & Values / Details   \\
\hline
Transducer Type         & L11-5v Linear array   \\
No. of elements ($N$)     & 128                   \\
Element pitch ($p$)           & 0.3 mm                \\
Center frequency  ($f_c$)      & 7.6 MHz               \\
Transmit type   & Non-steered plane waves   \\
Frame rate (PRF)    & 2000 frames per second (fps) \\
Sampling frequency ($f_s$)  & 31.25 MHz         \\
\hline \hline
\end{tabular*}
\label{ExpParams}
\end{table}

%\textcolor{red}{Insert a table containing the data acquisition/imaging parameters}
%%%%%%%%%%%%%%%%%%%

%%%%%%%%%%%%%%%%%
\begin{figure}[t]
\centering
\centerline{\includegraphics[width=0.3\textwidth]{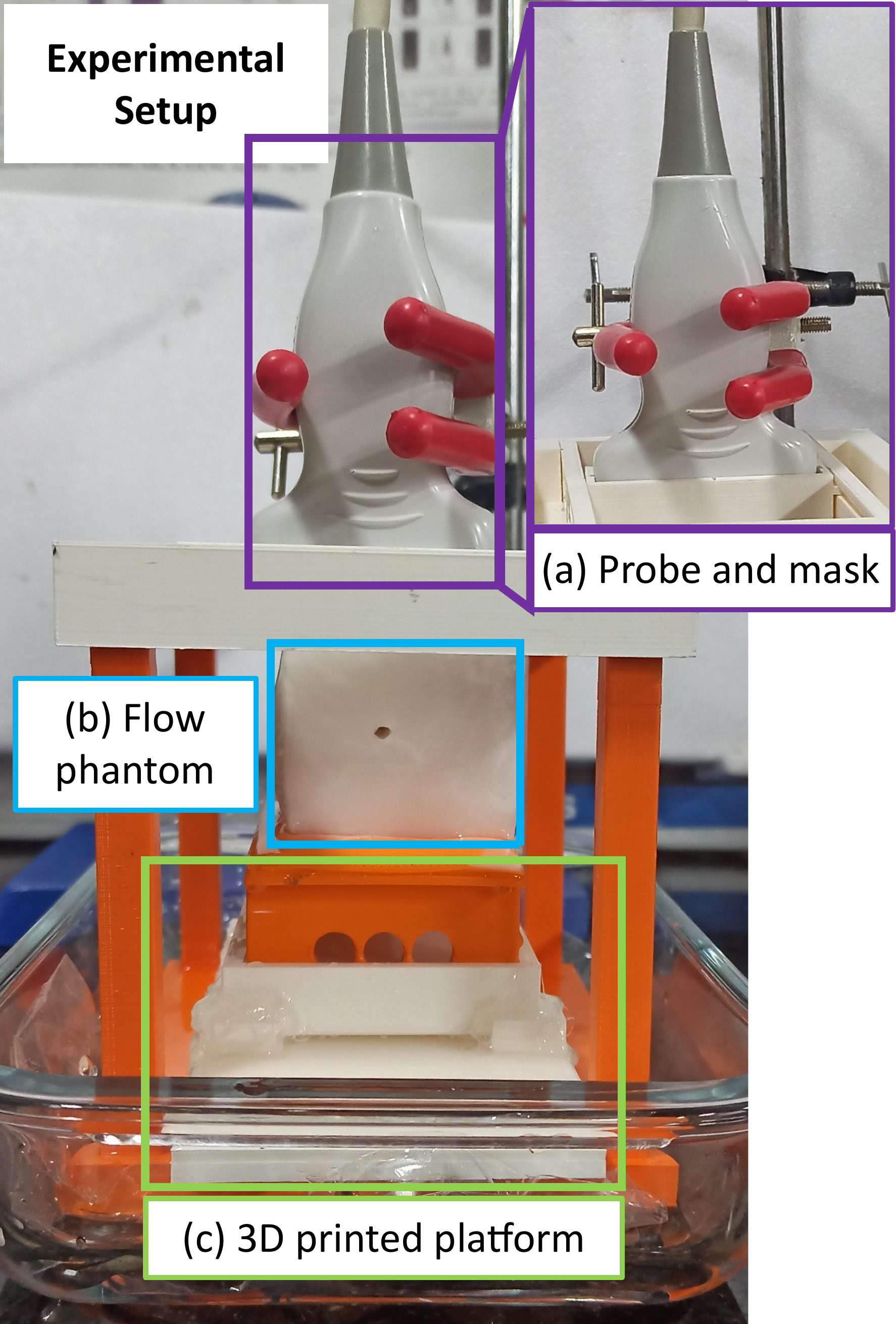}}
\caption{Experimental Setup: (a) Linear probe and probe mask, (b) PVA-based flow phantom, (c) 3D printed platform for phantom.}
\label{fig:exptSetup}
\end{figure}
%%%%%%%%

\subsubsection{Data Acquisition}
The BMF was pumped using a syringe into the channel in the flow phantom and raw datasets were acquired using Verasonics Vantage 128 Research Ultrasound System with a linear array of $7.6\ MHz$ center frequency. Non-steered plane waves at a pulse repetition frequency of $2000\ Hz$ were used to insonify the region of interest (ROI). The data acquisition parameters are shown in Table \ref{ExpParams} and the experimental setup is shown in Fig. \ref{fig:exptSetup}. A 3D printed probe mask (Fig. \ref{fig:exptSetup}(a)) and a raised platform (Fig. \ref{fig:exptSetup}(c)) to place the phantom (Fig. \ref{fig:exptSetup}(b)) were used to keep the scan plane intact during data acquisition and to ensure that the observations are not skewed due to any changes in the insonified region \cite{jerald2023simplified, jerald2023towards}.

%%%%
\begin{figure*}[t]
\centering
\centerline{\includegraphics[width=\textwidth]{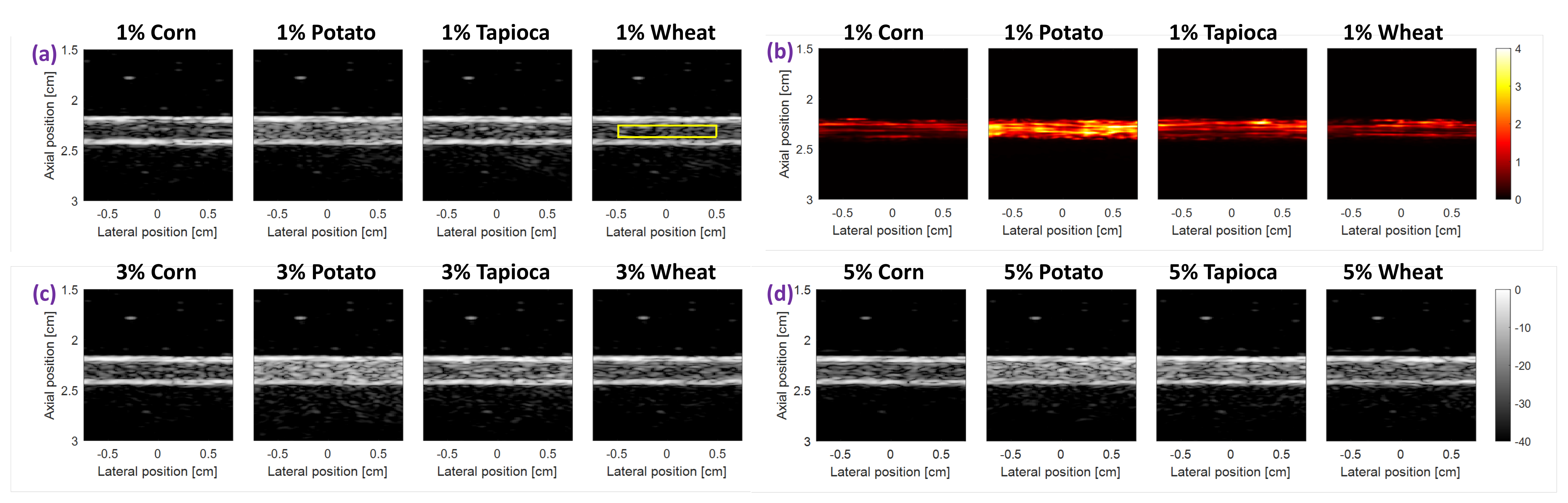}}
\caption{(a) B-mode image of PVA-based flow phantom with $1\%$ starch solution pumped (b) Power Doppler of PVA-based flow phantoms with $1\%$ starch solution pumped (c) B-mode image of PVA-based flow phantom with $3\%$ starch solution pumped (d) B-mode image of PVA based flow phantom with $5\%$ starch solution pumped. The region highlighted in yellow in (a) indicates the ROI chosen for the evaluation of echogenicity.}
\label{fig:BmodeImgs}
\end{figure*}
%%%%%
\begin{figure}[t]
\centering
\centerline{\includegraphics[width=0.49\textwidth]{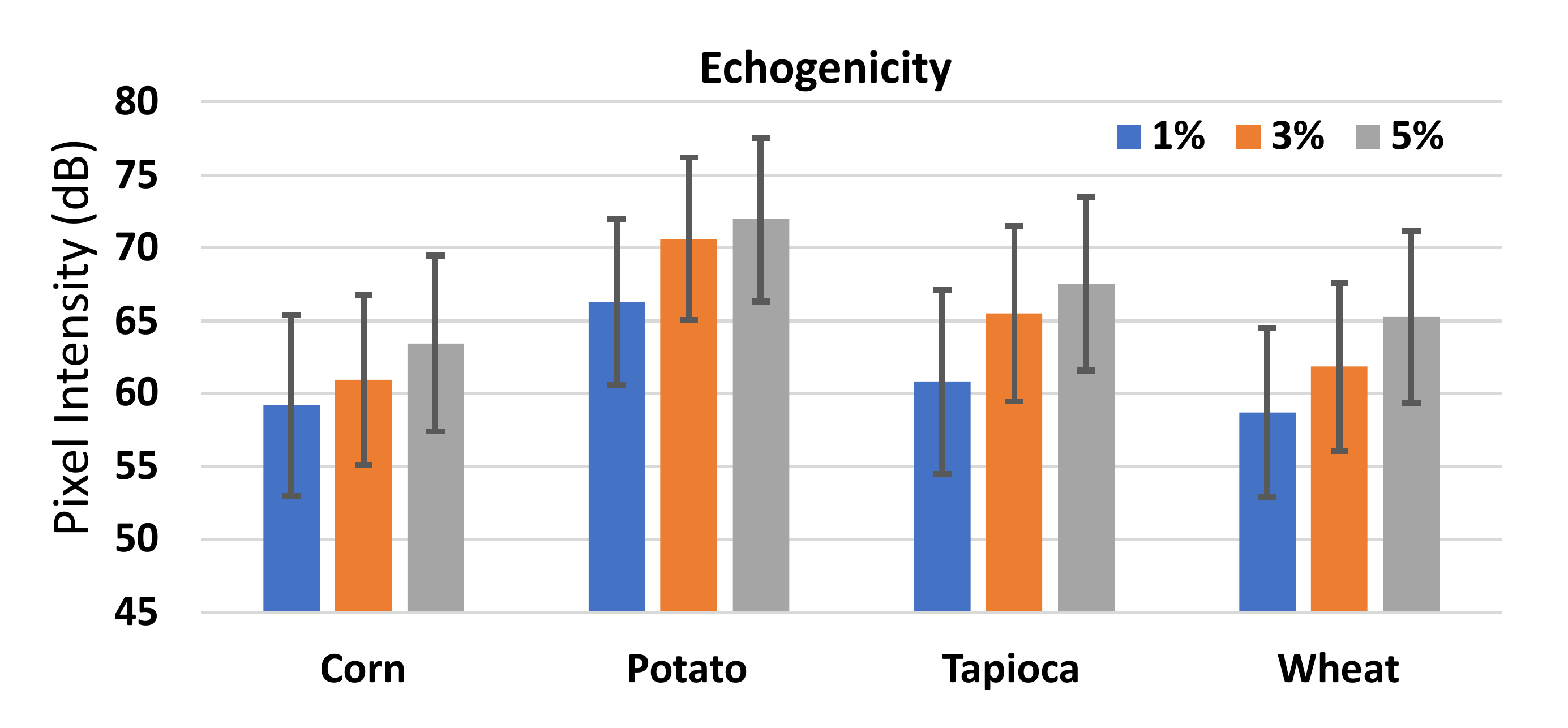}}
\caption{Plot of echogenicity of the starch-based BMFs with varying starch concentrations. The total length of error bars corresponds to $2\ \times$ standard deviations, indicating $\pm$standard deviation from the mean estimated intensity.}
\label{fig:statPlot}
\end{figure}
%%%%
%\subsection{Characterization of BMF}
% \begin{comment}
% \subsubsection{Density}
% A container was weighed using HSCo, Electronic Highly Precise Lab Scale($0.0001g$ precision) with and without a known volume($5 mL$) of samples and difference of mass was taken. This difference was then divided by the volume, obtaining the density. The measurement of density was carried out at $26^{\circ}$C.
% \end{comment}
% \begin{comment}
    
% \subsubsection{Viscosity}
% The dynamic viscosity of the samples was measured using a Modular Compact Rheometer(MCR) 702 (Anton Paar). \textcolor{blue}{To add after measuring.} 

% \textcolor{red}{This shall be omitted in this submission for 2 reasons: 1) The paper title is focused on the echogenicity measurement and not on the comprehensive characterization of BMF. 2) Since we are planning an extension of this work for a journal submission.}
% \end{comment}

\subsection{Echogenicity Measurement}
Echogenicity, by definition, refers to the capability of the tissue to produce echoes, when insonified with ultrasound waves, by reflecting or transmitting it with respect to the surrounding tissues \cite{ihnatsenka2010ultrasound}. Typically, it appears as a noticeable change in the contrast whenever there is an interface of regions with different acoustic properties. In practice, a tissue is characterized as anechoic, hypoechoic, and hyperechoic based on its echogenicity as apparent from the appearance of tissue (black, gray, and white respectively) in the B-mode image. %An object’s echogenicity allows it to be categorized as anechoic (black on the screen), hyperechoic (white on the screen), or hypoechoic (grey on the screen) \cite{ihnatsenka2010ultrasound}. 

In this work, conventional delay and sum (DAS) based beamforming was employed for image reconstruction. The echogenicity was measured by estimating the mean pixel intensity in a selected ROI in the beamformed image after log compression for 10 sets of 100 frames acquired at 2000 frames per second. The mean and standard deviation of echogenicity estimated for the ROI in 1000 frames are reported.
\begin{comment}
\subsubsection{Particle Size}
A smear of the BMF samples were prepared and was viewed under Optical Microscope at \textcolor{red}{50x} magnification. The images were obtained as shown in the Figure, and the average particle size was taken.\textcolor{blue}{To add after measuring.}
\end{comment}

% \begin{comment}
% \textcolor{blue}{The echogenicity of corn-, potato-, tapioca-, and wheat starch-based BMFs with three different starch concentrations (1\%, 3\%, and 5\%) were evaluated in the study. BMFs were prepared using a liquid base of pure water-glycerol mixture at a volume/volume ratio of 55\%-44\% for 1\% scratch, 53\%-44\% for 3\% scratch, and 51\%-44\% for 5\% starch concentrations. The BMFs were manually pumped to a PolyVinyl alcohol-based flow phantom using a syringe. The dataset was acquired using Verasonics Vantage 128 Research Ultrasound System with a linear array of 7.6 MHz center frequency. A 3D printed probe mask and a raised platform were used to keep the scan plane intact during data acquisition, as shown in the experimental setup in (a), and to ensure that the observations are not skewed due to any changes in the insonified region. The echogenicity was measured by estimating the mean pixel intensity in a selected region of interest (ROI) in the beamformed image before log compression for 10 sets of 100 frames acquired at 2000 frames per second.}
% \end{comment}

\section{Results and Discussion} \label{sec:results}
%\subsection{Particle Size}
%\subsection{Echogenicity}
Fig. \ref{fig:BmodeImgs}(a) and Fig. \ref{fig:BmodeImgs}(b) show the B-mode images and power Doppler images obtained for datasets of 1\% starch-based BMFs. The B-mode images obtained for datasets of 3\% and 5\% starch-based BMFs are shown in Fig. \ref{fig:BmodeImgs}(c) and (d) respectively. The ROI chosen for the evaluation of echogenicity is highlighted in yellow in Fig. \ref{fig:BmodeImgs}(a). From the B-mode images, it was observed that the contrast increased with the increase in starch concentration for all the BMFs. It was also observed that the contrast of the BMFs was different for different starches present in them with potato starch-based BMF being the most echoic while corn and wheat starch-based BMF being the least echoic. 

The qualitative results were validated quantitatively with echogenicity and the mean and standard deviation of estimated pixel intensities for the BMFs with different concentrations are shown in Fig. \ref{fig:statPlot} as pixel intensity in dB. It was observed that the intensity (in dB) of potato starch-based BMF was almost $13\%$, $14\%$, and $10\%$ higher than that of the least echoic BMF for $1\%$, $3\%$, and $5\%$ starch concentrations respectively. Moreover, the echogenicity of corn, tapioca, and wheat starch-based BMF was observed to be similar, and the results suggest that these BMFs with higher starch concentrations shall be employed for \textit{in-vitro} CEUS studies. However, further studies are required to characterize these BMFs and it will be taken up as a future work.
% The mean and standard deviation of the estimated echogenicity, normalized with 1\% corn starch values, are shown in Fig. \ref{fig:statPlot}. 

% \section{Conclusion} \label{sec:conclusion}
% A study comparing the echogenicity of four natural starch-based fluids has been conducted to investigate the potential of these starch-based fluids as BMFs for \textit{in-vitro} CEUS imaging. 
% \textcolor{blue}{Potato starch-based BMF has shown the highest echogenicity, with a \% increase of 13\%, 14\%, and 10\% higher than that of the least echoic BMF for 1\%, 3\%, and 5\% starch concentrations respectively.}

% Among the four natural starch-based BMFs considered in this study, potato starch-based BMF has shown the highest echogenicity and contrast as evident from Fig. \ref{fig:BmodeImgs} and \ref{fig:statPlot}. Moreover, the echogenicity of corn, tapioca, and wheat starch-based BMF was observed to be similar, and the results suggest that these BMFs with higher starch concentrations shall be employed for \textit{in-vitro} CEUS studies.

\section*{Acknowledgment}
We acknowledge the CSquare Innovation Lab, Central Micro-Nano Fabrication Facility, Physical and Chemical Biology Lab, and the Centre for Computational Imaging, Indian Institute of Technology Palakkad, Kerala for providing the required facilities for phantom development, ultrasound imaging, and BMF characterization. 
The authors are grateful to the Medical Ultrasound Engineering (MUSE) Lab, Indian Institute of Technology Gandhinagar, Gujarat for providing the PVA for phantom preparation and the technical discussions related to phantom preparation.
%Thanks to MUSE Laboratory IITGN for providing with PVA and for useful discussions.

\bibliographystyle{IEEEtran}
% \bibliography{strings,refs}
\small\bibliography{main_arXiv.bib}

\end{document}